\documentclass[preprint,5p]{elsarticle}

\usepackage{graphicx,amssymb}
\usepackage{epsf,psfig}

\journal{New Astronomy}

\begin{document}

\begin{frontmatter}

\title{Transition between order and chaos in a composite disk galaxy model with \\
a massive nucleus and a dark matter halo}

\author{Nicolaos D. Caranicolas}
\author{Euaggelos E. Zotos\corref{}}

\address{Department of Physics, \\
Section of Astrophysics, Astronomy and Mechanics, \\
Aristotle University of Thessaloniki \\
541 24, Thessaloniki, Greece}

\cortext[]{Corresponding author: \\
\textit{E-mail address}: evzotos@physics.auth.gr (Euaggelos E. Zotos)}

\begin{abstract}

We investigate the transition from regular to chaotic motion in a composite galaxy model with a disk-halo, a massive dense nucleus and a dark halo component. We obtain relationships connecting the critical value of the mass of the nucleus or the critical value of the angular momentum $L_{zc}$, with the mass $M_h$ of the dark halo, where the transition from regular motion to chaos occurs. We also present 3D diagrams connecting the mass of nucleus the energy and the percentage of stars that can show chaotic motion. The fraction of the chaotic orbits observed in the $(r,p_r)$ phase plane, as a function of the mass of the dark halo is also computed. We use a semi-numerical method, that is a combination of theoretical and numerical procedure. The theoretical results obtained using the version 8.0 of the Mathematica package, while all the numerical calculations were made using a Bulirsch-St\"{o}er FORTRAN routine in double precision. The results can be obtained in semi-numerical or numerical form and give good description for the connection of the physical quantities entering the model and the transition between regular and chaotic motion. We observe that the mass of the dark halo, the mass of the dense nucleus and the $L_z$ component of the angular momentum, are important physical quantities, as they are linked to the regular or chaotic character of orbits in disk galaxies described by the model. Our numerical experiments suggest, that the amount of the dark matter plays an important role in disk galaxies represented by the model, as the mass of the halo affects, not only the regular or chaotic nature of motion but it is also connected with the existence of the different families of regular orbits. Comparison of the present results with earlier work is also presented.

\end{abstract}


\end{frontmatter}

\section{Introduction}

Interesting information concerning the mass and the dynamical properties of a disk galaxy can be obtained from the rotation curve, that is the curve giving the rotational velocity $\Theta(r)$ as a function of the radius $r$. It is well known, that disk galaxies do not rotate like a solid body. Stars and gas, in these systems, move in nearly circular orbits, while the angular velocity of rotation typically decreases with radius. The total mass $M(r)$ within radius $r$ can be found assuming circular motion, a mass point model for the galaxy and equating centrifugal and gravitational force obtaining
\begin{equation}
M(r) = \frac{r \Theta ^2 (r)}{G},
\end{equation}
where $G$ is the gravity constant. For a constant value of the mass one expects the rotational velocity to decline at a rate $\Theta(r) \propto r^{-1/2}$. Observations show that instead of declining the rotation curves, in most disk galaxies, remain flat at large radii. This implies that $M(r) \propto r$, meaning that the mass of the galaxy continues to grow, even when there is no visible matter to account for this increase. This is a clear indication that there must be much more matter in a disk galaxy that can be accounted for by the visible light we see. The first clear evidence for the presence of dark matter in galaxies was given by Rubin \& Ford (1970).

Observational data indicate that disk galaxies are often surrounded by haloes. Galactic haloes can have a variety of shapes from spherical to flattened and triaxial (see e.g. Dubinski \& Carlberg, 1991; Cooray, 2000; Kunihito et al., 2000; Olling \& Merrifield, 2000; Yoshida et al., 2000; Avila-Reese et al., 2001; Jenkins et al., 2001; Klypin et al., 2001; Jing \& Suto, 2002; Wechsler et al., 2002; Kasun \& Evrard, 2005; Allgood et al., 2006; Papadopoulos \& Caranicolas 2006; Capuzzo-Dolcetta et al., 2007; Wang et al., 2009; Evans et al., 2009; Caranicolas \& Zotos, 2010; 2011). In order to help the reader to have a better view of the galactic haloes we must make clear the following: in the stellar halo, which is the visible part of the halo, one observes population II objects, including globular clusters and old individual stars. Beyond this visible part, there is the dark halo, an extended region containing large amounts of dark matter. We must emphasize, that the stellar halo is not the inner part of the dark halo. They are two physically distinct components with different formation histories (see Caranicolas \& Zotos, 2009). However, the exact nature of dark matter in the galactic haloes of disk galaxies is still undetermined.

As most disk galaxies are surrounded by massive dark halo components, there is no doubt that the behavior of the orbits in the galactic disk is affected by the presence of the dark halo. Important role for the regular or chaotic nature of motion is played by the mass of the dark halo. In order to investigate the character of motion in disk galaxies we adopt a composite mass model with a disk-halo, a dense nucleus, and a spherical dark halo component. The total potential $V_t$ for our composite model, in cylindrical coordinates $(r, \phi, z)$, is given by the equation
\begin{equation}
V_t(r,z) = V_d(r,z) + V_n(r,z) + V_h(r,z),
\end{equation}
where
\begin{equation}
V_d(r,z) = - \frac{M_d}{\sqrt{b^2 + r^2 + \left(\alpha + \sqrt{h^2 + z^2}\right)^2}},
\end{equation}
\begin{equation}
V_n(r,z) = - \frac{M_n}{\sqrt{r^2 + z^2 + c_n^2}},
\end{equation}
while
\begin{equation}
V_h(r,z) = - \frac{M_h}{\sqrt{r^2 + z^2 + c_h^2}}.
\end{equation}
The first component, Eq. (3), is a disk-halo potential. $M_d$ is the mass of the disk, while $\alpha$, $h$ and $b$ correspond to the disk's scale length, the disk's scale height and the core radius of the disk-halo respectively, (see Carlberg \& Innanen, 1987; Caranicolas and Innanen, 1991). The second component part, Eq. (4), is the potential of a spherical dense nucleus. Here, $M_n$ and $c_n$ is the mass of and the scale length of the nucleus respectively. Finally, the third component Eq. (5), is the potential of a spherical dark halo of mass $M_h$ and a core radius $c_h$.

The aim of the present paper is, among others: (i) To investigate the transition from regular motion to chaos in the composite model (2) and to present relationships connecting: (a) the critical value of the mass of the nucleus - that is the minimum value of mass of nucleus required for the transition from regular to chaotic motion and the mass of halo, (b) the critical value of angular momentum - that is the maximum value of the angular momentum required for the transition from regular to chaotic motion and the mass of halo. In both cases, all other parameters are kept constant. (ii) To present three-dimensional (3D) plots connecting: (a) the energy, the mass of nucleus and the percentage of stars that can show chaotic motion and (b) To present 3D relationships between the Lyapunov Characteristic Exponent (LCE) (see Lichtenberg \& Liebermann, 1992), the percentage of chaotic regions on the phase plane and the angular momentum.

In our numerical calculations we use a system of galactic units, where the unit of length is 1 kpc, the unit of mass is 2.325 $\times$ $10^7$ M$_{\odot}$ and the unit of time is 0.97748 $\times$ $10^8$ yr. The velocity unit is 10 km/s, while $G$ is equal to unit. For the orbit calculations we have used a Bulirsch-St\"{o}er FORTRAN routine in double precision and the accuracy of our calculations was checked by the constancy of the energy integral, which was conserved up to the fifteenth significant decimal point. For all other numerical calculations we used the version 8.0 of the Mathematica package. The values of the parameters used are: $M_d = 8000$, $\alpha = 3$, $b = 8$, $h = 0.18$, $c_h = 12$, $c_n = 0.25$, while $L_z$, $M_h$ and $M_n$ are treated as parameters.
\begin{figure}[!tH]
\centering
\resizebox{\hsize}{!}{\rotatebox{0}{\includegraphics*{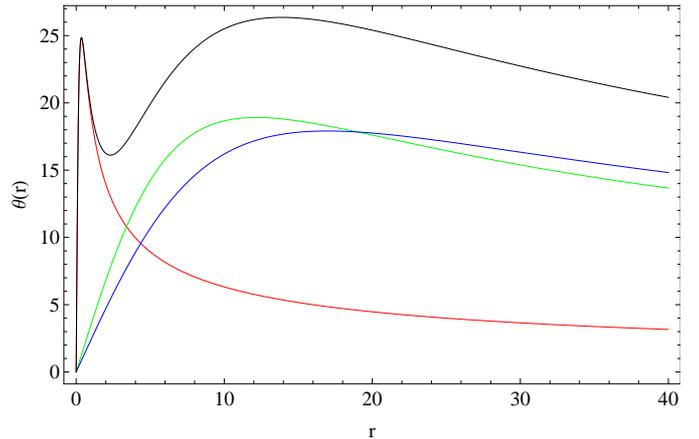}}}
\caption{The rotation curve for the disk galaxy model (2), when $M_n = 400$ and $M_h = 10^4$. The black line shows the total circular velocity. The red line is the contribution from the spherical nucleus, the green line comes from the disk, while the blue line represents the contribution from the spherical dark halo component. The values of all the other parameters are: $M_d = 8000$, $\alpha = 3$, $b = 8$, $h = 0.18$, $c_h = 12$ and $c_n = 0.25$.}
\end{figure}

Figure 1 shows the rotation curve for the disk galaxy model (2) derived using the formula
\begin{equation}
\Theta(r) = \sqrt{r\frac{\partial V_t(r,0)}{\partial r}}.
\end{equation}
The values of all the other parameters are as above, while $M_n = 400$ and $M_h = 10^4$. The total circular velocity is shown as the black line. The red line is the contribution from the spherical nucleus, the green line stands for the contribution from the disk, while the blue line represents the contribution from the spherical dark halo component.

The paper is organized as follows: in Section 2 we present the numerically found new relationships and we try to express them using simple mathematical expressions. In Section 3 we study the structure of the phase plane in some representative cases and present the different families of orbits. We conclude with a discussion and a comparison with earlier work, which is presented in Section 4.

\section{Linking physical parameters to chaos}

It was more than twenty years ago, when Caranicolas \& Innanen, (1991) found, using a semi-numerical procedure, a linear relationship between the critical value of the star's angular momentum $L_{zc}$ and the mass of nucleus in a disk galaxy model with a dense and massive nucleus. Numerical calculations show that this linear relationship still holds, when the galaxy model is surrounded by a dark halo of mass $M_h$, that is, in the case of the model (2). Figure 2 shows the obtained relationship between the critical value $L_{zc}$ and $M_n$, when $M_h =10^4$. The values of all the other parameters are: $M_d = 8000$, $\alpha = 3$, $b = 8$, $h = 0.18$, $c_h = 12$ and $c_n = 0.25$. Our numerical calculations indicate that the relationship between $L_{zc}$ and $M_n$ is nearly a straight line which is presented in Fig. 2. This straight line divides the $(L_z, M_n)$ plane in two parts. Orbits with starting conditions on the left shaded part (which also contains the line) are chaotic, while orbits with starting conditions on the right part of the same diagram are regular. Note, that slope of the line is different for very low values of $L_z$. In order to help the reader, we must remind that the critical value $L_{zc}$ is the maximum value of the angular momentum $L_z$ required to observe chaos, for a given value of mass of the nucleus, when all other parameters are kept constant. Thus we write
\begin{equation}
M_n \propto L_{zc}.
\end{equation}
\begin{figure}[!tH]
\centering
\resizebox{\hsize}{!}{\rotatebox{0}{\includegraphics*{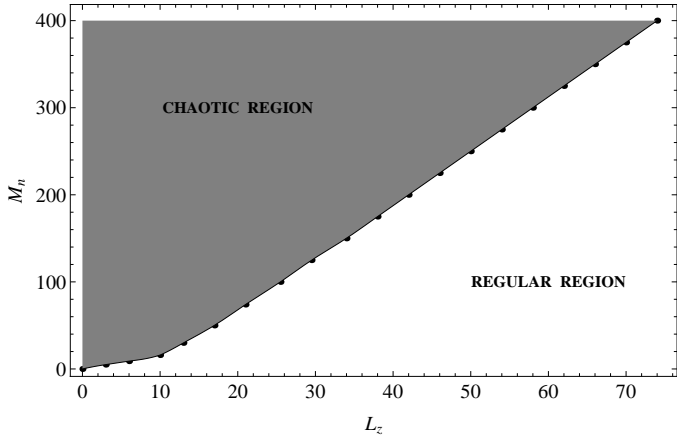}}}
\caption{Relationship between the critical value $L_{zc}$ and $M_n$, when $M_h = 10^4$. The values of all the other parameters are as in Fig. 1.}
\end{figure}
\begin{figure}[!tH]
\centering
\resizebox{\hsize}{!}{\rotatebox{0}{\includegraphics*{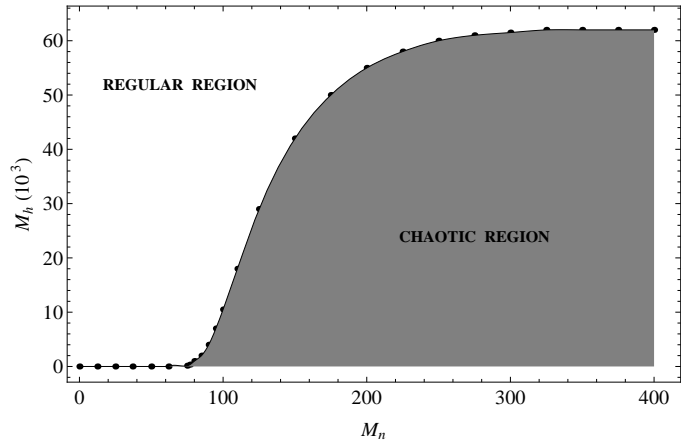}}}
\caption{A plot of $M_h$ versus the critical value of $M_n$, when $L_z = 25$. The values of all the other parameters are as in Fig. 1.}
\end{figure}
\begin{figure}[!tH]
\centering
\resizebox{\hsize}{!}{\rotatebox{0}{\includegraphics*{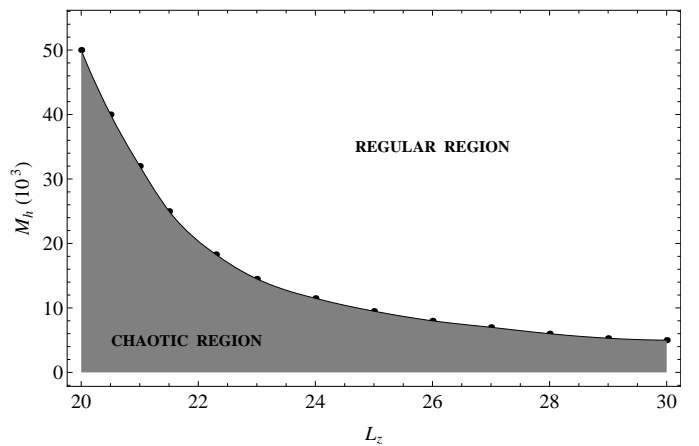}}}
\caption{A plot $M_h$ versus the critical value $L_{zc}$, when $M_n = 100$. The values of all the other parameters are as in Fig. 1.}
\end{figure}

Besides the well known relationship (7), our numerical experiments gave interesting links between the mass of the halo $M_h$ and the critical value of angular momentum, as well as, links between the mass of the dark halo $M_h$ and the critical value of the mass of the nucleus, i.e the minimum mass of the nucleus required to observe chaos, when all other parameters remain constant. Figure 3 shows a plot of $M_h$ versus the critical value of $M_n$, when $L_z = 25$. It is interesting to observe that for small values of the mass of the nucleus, we have $M_h = 0$, while for large values of the mass of the nucleus the curve remains nearly flat, as $M_n$ increases. The values of all the other parameters are as in Fig. 2. Figure 4 shows a plot of $M_h$ versus the critical value $L_{zc}$, when $M_n = 100$. As one can see, $M_h$ increases non-linearly as the critical value $L_{zc}$ decreases. Thus we see, that apart from the linear relationship (7), we have found numerically two more non-linear relationships described by Figs. 3 and 4. All the above relationships can be described by the general expression
\begin{equation}
M_n \propto L_{zc} M_h^2.
\end{equation}
Relationship (8) is a simple formula connecting three important galactic physical quantities the mass of the nucleus, the critical value $L_{zc}$ and the mass $M_h$ of the dark halo. Note that expression (8) is reduced to relationship (7), when $M_h = const$. It is interesting to note, that expression (8) suggests that $L_{zc}$ decreases nonlinearly as $M_h$ increases, when $M_n$ is kept constant and also indicates that $M_h$ increases nonlinearly, as $M_n$ increases, when $L_{zc}$ is kept constant. Here, we must make clear that relationship (8) is a simple empirical formula, which is based in our experience, describing satisfactory the nonlinear dependence between the physical quantities entering Fig. 3 and Fig. 4. The reader must always have in mind that neither expression (7) nor formula (8) can reproduce the numerically found results. On the other hand, it is very useful because it can give us interesting information regarding the connection of some important  physical quantities of the galaxy described by the model (2) and the transition from regular motion to chaos. Also note that relationship (8) does not reproduce the straight line segments of Fig. 3.
\begin{figure}[!tH]
\centering
\resizebox{\hsize}{!}{\rotatebox{0}{\includegraphics*{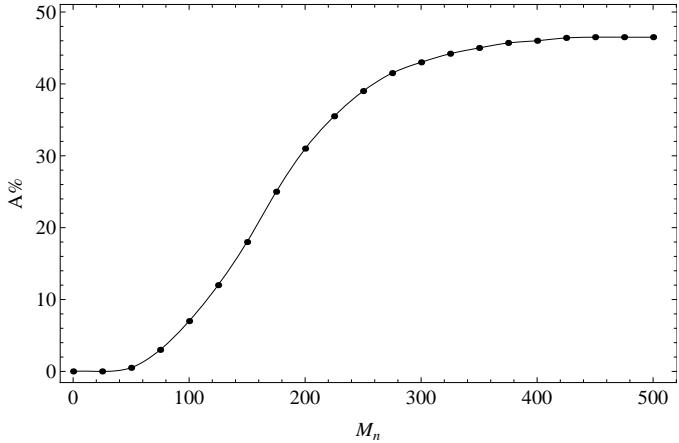}}}
\caption{A plot of the percentage $A\%$ versus $M_n$, when $L_z = 20$ and $M_h = 10^4$. The values of all the other parameters are as in Fig. 1.}
\end{figure}

It is of particular interest to investigate how the physical parameters of the model are connected with the percentage $A\%$ of chaotic orbits. The percentage $A\%$ is found in a completely empirical way, by measuring the area on the $(r, p_r)$ phase plane covered by chaotic orbits. Figure 5 shows a plot of $A\%$ versus $M_n$, when $L_z = 20$ and $M_h = 10^4$. As one can see, for small values of $M_n$, the chaotic regions are small, while for massive nuclei almost half of the phase plane is covered by chaotic orbits. Figure 6 shows the $A\%$ as a function of $M_h$, when $M_n = 100$ and $L_z = 20$. We observe that $A\%$ decreases nonlinearly as $M_h$ increases. Figure 7 shows a 3D plot connecting the value of the angular momentum $L_z$, the LCE and the percentage $A\%$, when $M_n = 100$ and $M_h = 10^4$. Note that, for large values of $L_z$ no chaotic motion is observed, while both LCE and $A\%$ increase as $L_z$ decreases.
\begin{figure}[!tH]
\centering
\resizebox{\hsize}{!}{\rotatebox{0}{\includegraphics*{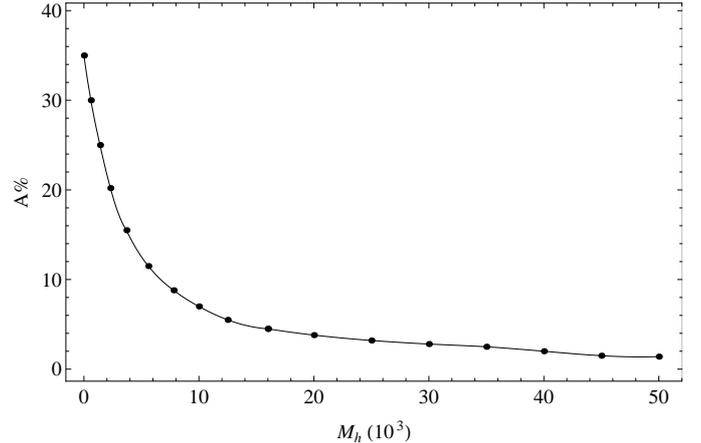}}}
\caption{A plot of the percentage $A\%$ as a function of $M_h$, when $M_n = 100$ and $L_z = 20$. The values of all the other parameters are as in Fig. 1.}
\end{figure}
\begin{figure}[!tH]
\centering
\resizebox{\hsize}{!}{\rotatebox{0}{\includegraphics*{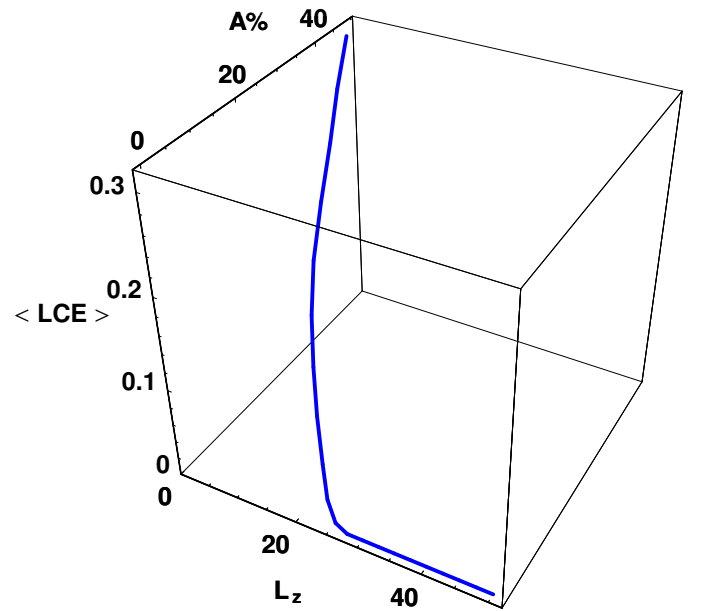}}}
\caption{A 3D plot connecting the value of the angular momentum $L_z$, the LCE and $A\%$, when $M_n = 100$ and $M_h = 10^4$. The values of all the other parameters are as in Fig. 1.}
\end{figure}
\begin{figure}[!tH]
\centering
\resizebox{\hsize}{!}{\rotatebox{0}{\includegraphics*{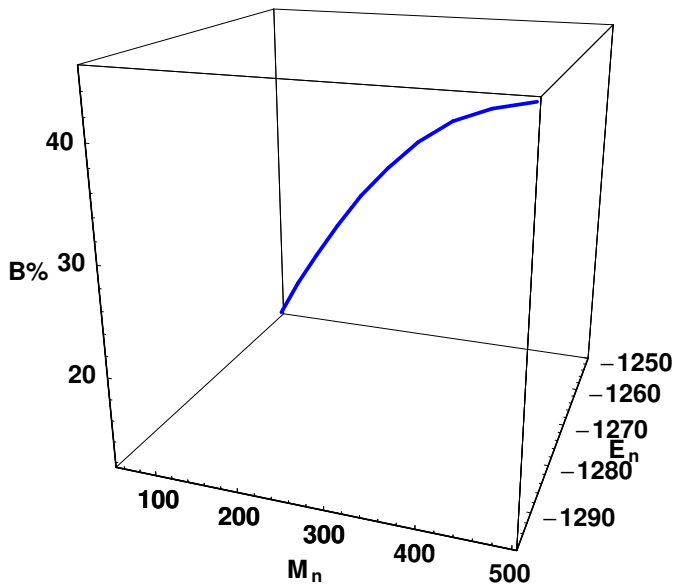}}}
\caption{A 3D plot connecting $E_n$, $M_n$ and $B\%$.}
\end{figure}

One can derive useful information for the dynamical behavior of the galaxy described by the model from Figure 8. This Figure presents a 3D plot connecting the value of energy $E_n$, the mass of nucleus $M_n$ and the percentage of stars $B\%$ that can display chaotic motion, when they approach the massive nucleus. The value of the mass for the dark halo is fixed to $M_h = 10^4$. In order to construct this plot we work as follows: For a given value of the energy and the mass of the nucleus, let's say $E_n$ and $M_n$, we find the value of the effective potential at $z = 0$ (see next Section), which is given by
\begin{equation}
V_{eff}(r,0) = \frac{L_z^2}{2r^2} + V_t(r,0).
\end{equation}
Then we solve numerically equation (9) for $r_{max}$ and $r_{min}$ using the value of $L_z \neq 0$, as a parameter, starting from a small value of $L_z$, say $L_{zmin} = 0.01$. As the value of $L_z$ increases, $r_{min}$ and $r_{max}$ come close to each other and there is a value of $L_z$, where $r_{min} = r_{max}$ and this is the maximum value of $L_z$ for the stars with the above values of $E_n$ and $M_n$. Let $L_{zmax}$ be the maximum value of $L_z$. Our next step is to find the critical value of the angular momentum for the adopted values of $E_n$ and $M_n$. Let $L_{cr}$ be this value of $L_z$. Then we define
\begin{equation}
B = \frac{L_{cr} - L_{zmin}}{L_{zmax} - L_{zmin}}.
\end{equation}
Repeating the above procedure, for different couples of the values $(E_n, M_n)$, we construct the diagram given in Fig. 8. This is a 3D plot connecting $E_n$, $M_n$ and $B\%$. According to this diagram, the percentage of stars $B\%$, that can show chaotic motion, displays a nonlinear dependence on the energy and the mass of the nucleus and can reach about the $45\%$ of the total amount of the stars.

In this Section, we presented a new relationship connecting three basic physical quantities of the galaxy described by the new model (2). Furthermore, we constructed diagrams connecting the percentage of chaos $A\%$ with the mass of nucleus $M_n$ as well as $A\%$ and $M_h$. In addition to the above diagrams we presented two 3D plots. The first connecting $L_z$, LCE and $A\%$, while the second gives a relationship between $E_n$, $M_n$ and the percentage $B\%$ of stars that can display chaotic motion.
\begin{figure*}[!tH]
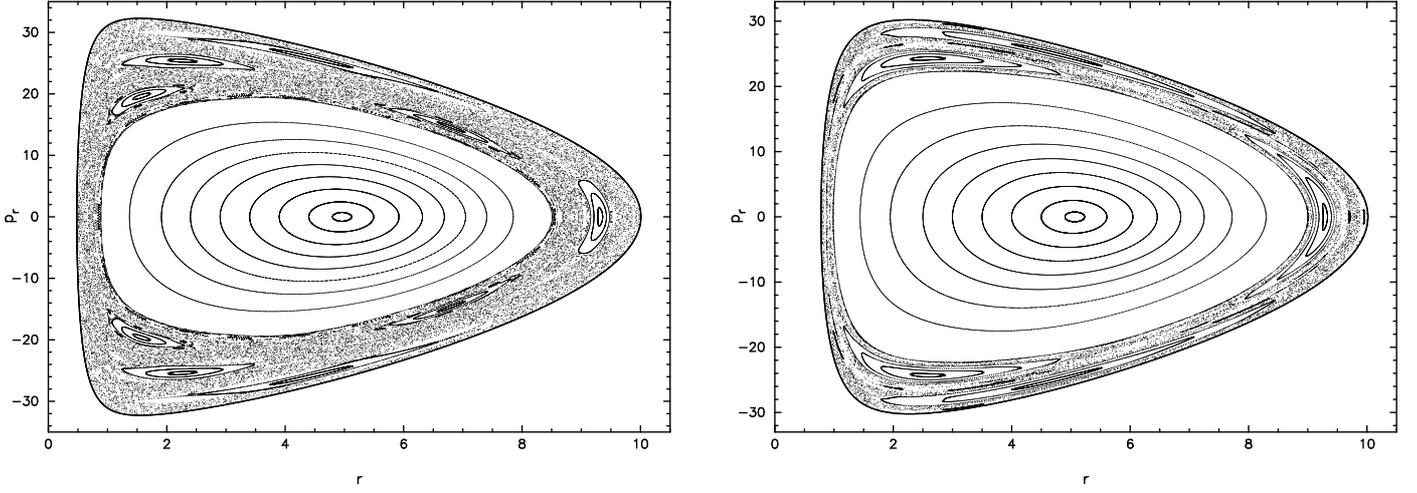

\centering
\resizebox{\hsize}{!}{\rotatebox{270}{\includegraphics*{Fig-9a.ps}}\hspace{2cm}
                      \rotatebox{270}{\includegraphics*{Fig-9b.ps}}}
\vskip 0.01cm
\caption{(a-b): The $(r, p_r)$ phase plane, when $M_n = 200$ and $M_h = 10^4$. (a-left): $L_z = 20$ and $E_n = -1264$. (b-right): $L_z = 30$ and $E_n = -1261$.}
\end{figure*}
\begin{figure*}[!tH]
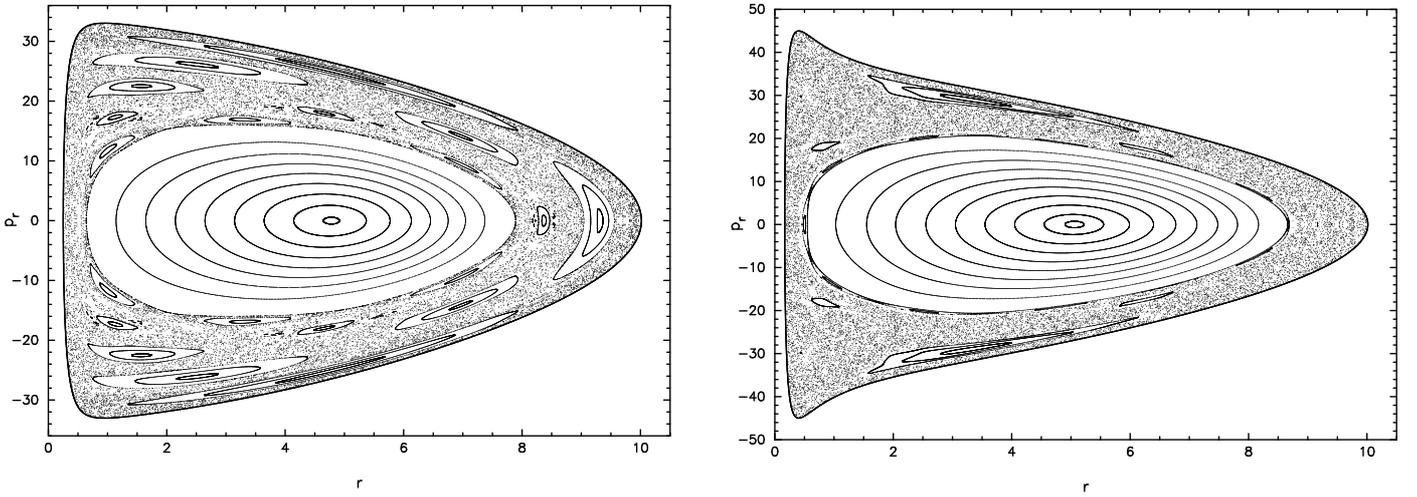

\centering
\resizebox{\hsize}{!}{\rotatebox{270}{\includegraphics*{Fig-10a.ps}}\hspace{2cm}
                      \rotatebox{270}{\includegraphics*{Fig-10b.ps}}}
\vskip 0.01cm
\caption{(a-b): Similar to Fig. 9a-b, when $M_h = 10^4$ and $L_z = 10$. (a-left): $M_n = 100$ and $E_n = -1255$. (b-right): $M_n = 400$ and $E_n = -1285$.}
\end{figure*}
\begin{figure*}[!tH]
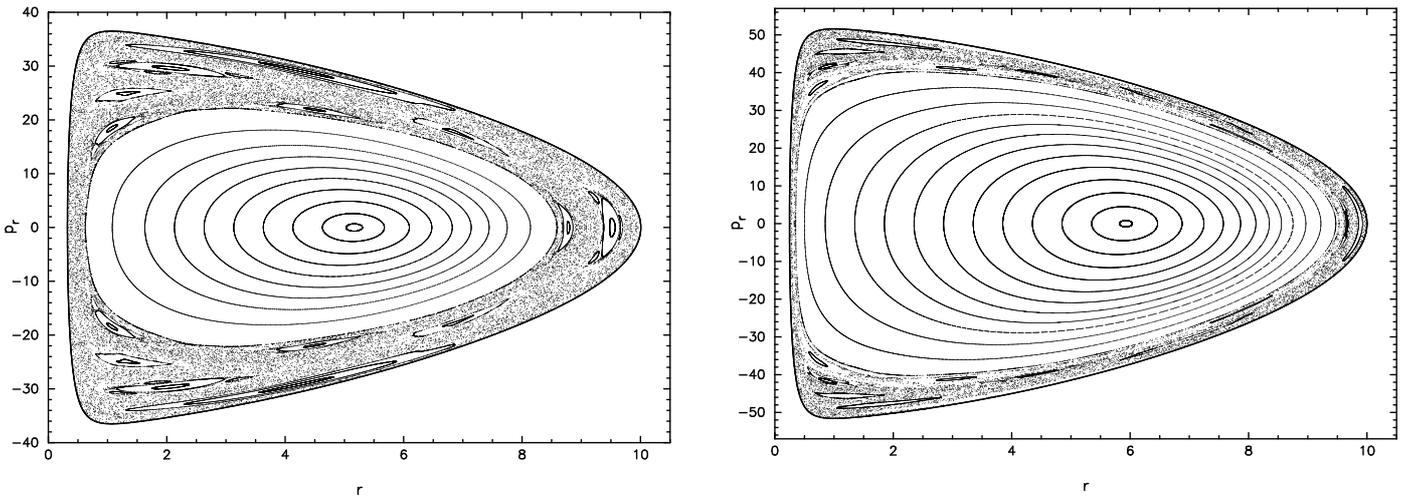

\centering
\resizebox{\hsize}{!}{\rotatebox{270}{\includegraphics*{Fig-11a.ps}}\hspace{2cm}
                      \rotatebox{270}{\includegraphics*{Fig-11b.ps}}}
\vskip 0.01cm
\caption{(a-b): Similar to Fig. 9a-b, when $M_n = 200$ and $L_z = 15$. (a-left): $M_h = 1.5 \times 10^4$ and $E_n = -1585$. (b-right): $M_h = 4 \times 10^4$ and $E_n = -3826$.}
\end{figure*}

\section{Families of regular orbits and chaos}

In this Section, we shall investigate the different families of regular orbits appearing in the new galaxy model and how these families are connected with the physical parameters of the system that is $L_z$, $M_n$ and $M_h$. Furthermore, we shall study,  if there is only one or several different chaotic components (see Saito \& Ichimura, 1979), in the galaxy described by the model (2).

In axially symmetric models, where the $L_z$ component of the star's angular momentum is conserved, we usually take the effective potential
\begin{equation}
V_{eff}(r,z) = \frac{L_z^2}{2r^2} + V_t(r,z),
\end{equation}
studying the motion in the meridian $(r, z)$ plane. The equations of motion are
\begin{eqnarray}
\dot{r} = p_r, \ \ \dot{z} = p_z, \nonumber \\
\dot{p_r} = - \frac{\partial V_{eff}}{\partial r}, \ \ \dot{p_z} = - \frac{\partial V_{eff}}{\partial z},
\end{eqnarray}
where the dot indicates derivative with respect to the time. The Hamiltonian corresponding to the effective potential (11) is
\begin{equation}
H = \frac{1}{2} \left(p_r^2 + p_z^2 \right) + V_{eff}(r,z) = E_n,
\end{equation}
where $p_r$ and $p_z$ are the momenta, per unit mass conjugate to $r$ and $z$ respectively, while $E_n$ is the numerical value of the Hamiltonian, which is conserved.
\begin{figure*}[!tH]
\centering
\resizebox{0.90\hsize}{!}{\rotatebox{0}{\includegraphics*{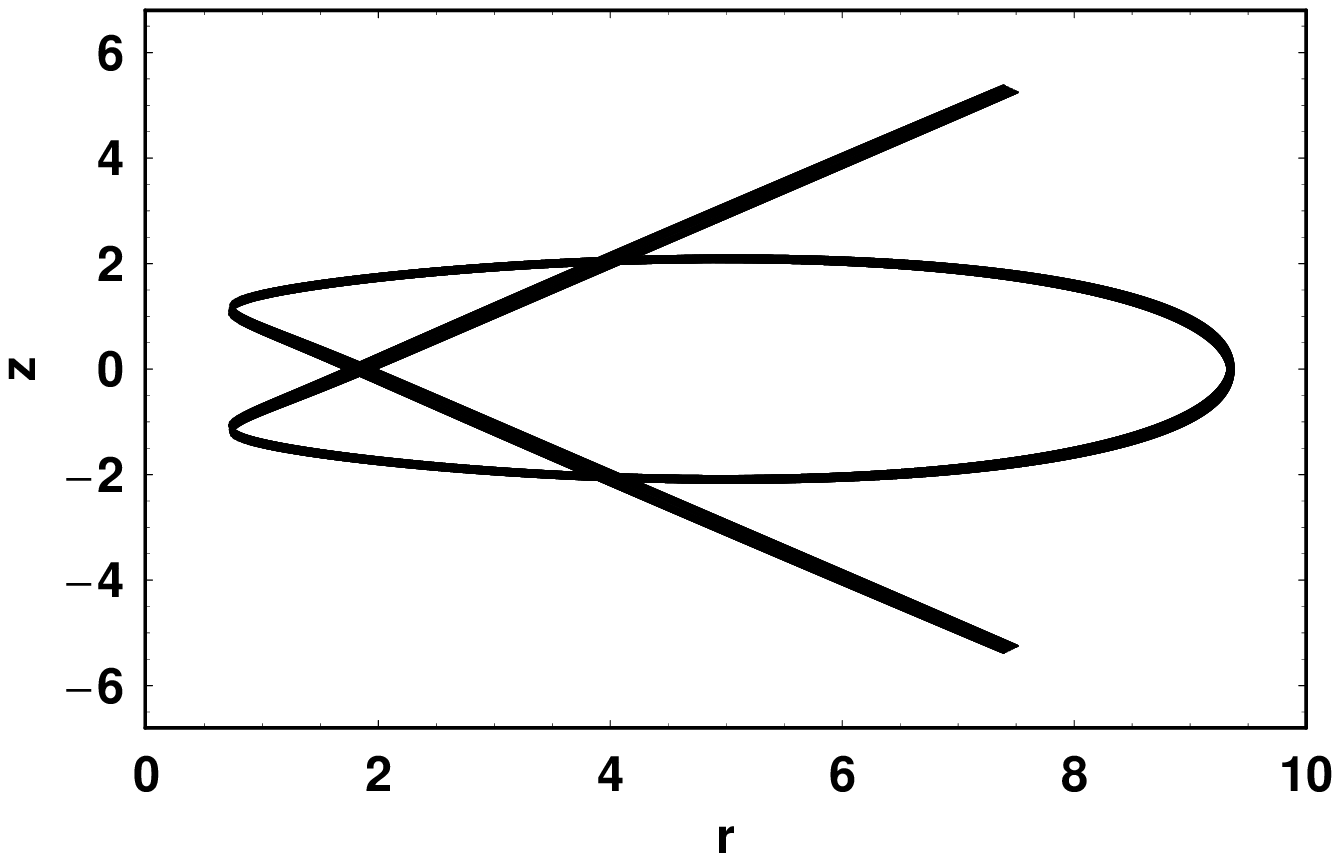}}\hspace{1cm}
                          \rotatebox{0}{\includegraphics*{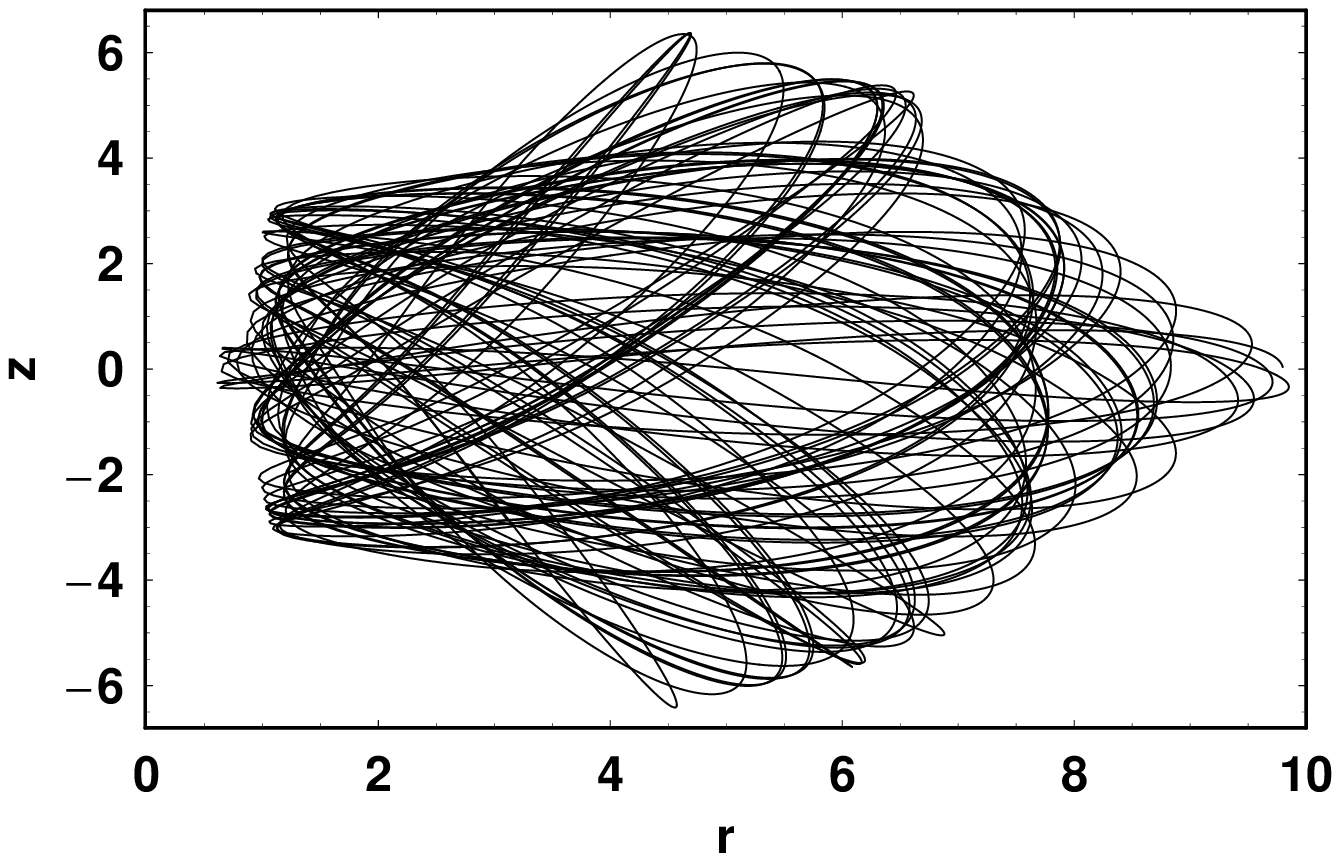}}}
\vskip 0.01cm
\caption{(a-b): (a-left): A regular orbit characteristic of the 4:3 resonance and (b-right): a chaotic orbit with the same initial conditions as in (a) but when $M_h = 0$. See text for more details.}
\end{figure*}

In the following we shall use the numerical integration of the equations of motion (12), in order to compute the Poincar\'{e} $(r, p_r)$, $z = 0$, $p_z > 0$ phase plane. As the method is classical, it gives interesting and reliable information on the regular or chaotic character of orbits and indications for the existence of the different families of orbits in the model. In all the numerical calculations we choose the value of energy $E_n$, such as to obtain $r_{max}$ about 10 kpc, when $z = 0$, while the values of the parameters, that are kept fixed, are: $M_d = 8000$, $\alpha = 3$, $ = 8$ $h = 0.18$, $c_h = 12$ and $c_n = 0.25$.

Figure 9a-b shows the $(r, p_r)$ phase plane, when $M_n = 200$ and $M_h = 10^4$. In Fig. 9a, where $L_z = 20$ and $E_n = -1264$, we observe regular motion together with a large unified chaotic sea. One can see the following basic families of regular orbits: (i) orbits producing invariant curves surrounding the central invariant point. These orbits are box orbits, (ii) orbits producing the set of elongated islands of invariant curves shown on the outer parts of Fig. 9a. Note that each orbit of the family produces a single island. These orbits are banana like orbits and are characteristics of the 1:1 resonance. (iii) orbits producing a set of two asymmetric islands of invariant curves. These orbits are loop orbits and (iv) orbits producing a set of three islands of invariant curves, one on the $r$ axis and the other two are symmetric with respect to the $r$ axis. These orbits are characteristics of the 4:3 resonance. The outermost solid line is the Zero Velocity Curve (ZVC). In Fig. 9b we have $L_z = 30$ and $E_n = -1261$. Here the chaotic layer is smaller, as the value of $L_z$ is larger, while the orbits of family (ii) are not present.

Figure 10a-b is similar to Fig. 9a-b, when $M_h = 10^4$ and $L_z = 10$. In Fig. 10a where $M_n = 100$ and $E_n = -1255$, we  can find the same four families of regular orbits as in Fig. 9a. Apart from the above families, we see sets of small islands produced by secondary resonances. All these small islands of invariant curves are embedded in a large chaotic sea. Things appear different in Figure 10b, where $M_n = 400$ and $E_n = -1285$. Here the chaotic sea is larger, because of the more massive nucleus. The regular families (i) and (ii) are also present, while family (iii) is now absent. Some tiny islands of invariant curves produced by secondary resonances are also observed. The role of a massive and dense nucleus in an axially symmetric galactic model and how it affects the ordered or chaotic nature of the orbits, was also investigated in Zotos (2012).

In the phase plane shown in Figure 11a-b we have adopted the values $M_n = 200$ and $L_z = 15$. In Fig. 11a, where $M_h = 1.5 \times 10^4$ and $E_n = -1585$, the structure of the phase plane appears similar to that of Fig. 10a. In Figure 11b, where $M_h = 4 \times 10^4$ and $E_n = -3826$, one can see the same families of regular orbits as in Fig. 11a but, in this case, the majority of the phase plane is covered by orbits belonging to family (i), that is box orbits. The remaining three families are also present but they occupy a small fraction of the phase plane. Small fraction of the phase plane is also covered by chaotic orbits.

Before closing this Section, we would like to present an illuminating example which points out the influence of the dark halo component on the nature of the orbits. Figure 12a-b presents two orbits. In Fig. 12a we observe a regular orbit characteristic of the 4:3 resonance. The values of the parameters are: $M_n = 400$, $M_h = 10^4$, $L_z = 25$ and $E_n = -1283$. The initial conditions are: $r_0 = 9.3$ and $z_0 = p_{r0} = 0$. On the other hand, in the orbit shown in Fig. 12b the value of $M_h$ is equal to zero, $E_n = -643$, while all the other parameters and the initial conditions are the same as in Fig. 12a. Therefore, we may conclude, that when the spherical dark halo component is not present, the orbit alters its character from regular to chaotic.

In this Section, we have studied the structure of motion in the new galaxy  model (2). Four basic families of regular orbits were found. In the case of chaotic motion, only one chaotic component was found. Secondary resonances were also present. An interesting result observed during the orbit calculations is that when a massive halo with a mass $M_h = 4 \times 10^4$ is present, the majority of orbits are box orbits, while the chaotic region is small, when $M_n = 200$ and $L_z = 15$. This suggests, that in galaxies with massive nuclei the majority of low angular momentum stars are in box orbits, when the galaxy is surrounded by a massive dark halo component.

\section{Discussion}

In this paper, we have studied the regular and chaotic nature of motion in a composite disk galaxy model with a massive nucleus and a dark halo component. Numerical calculations have shown that the transition from regular to chaotic motion is connected with three basic parameters of the system, that is the angular momentum $L_z$ the mass of the nucleus $M_n$ and the mass of  the dark halo $M_h$. In order to present a new interesting outcome, we have tried to express the numerically found results using a single relationship expression connecting the above three important parameters in the interesting case, where the transition from regularity to chaos occurs. As it was mentioned, this relationship indicates the linear or the nonlinear dependence of each couple of the three parameters $L_z$, $M_n$ and $M_h$, when the third is kept constant, for fixed values all the other parameters entering the model (2). Once again, we must say that relationship (8) is based in our experience. Note that the nonlinear dependence was expressed with an exponent 2 in the mass of the dark halo $M_h$, that is the smaller integer. This was made not only for simplicity but also because this exponent was suitable to express the form of the curves found numerically.

Interesting new results were also presented by the two 3D diagrams. In the first, connecting $L_z$, LCE and $A\%$, it was found that, both the LCE and the $A\%$, increase as the value of the $L_z$ decreases. In the second diagram we gave a relationship between $E_n$, $M_n$ and the percentage $B\%$ of stars that can display chaotic motion. It was found that $B\%$ can be as high as about the $45\%$ of the total amount of stars.

The study of the structure of the $(r, p_r)$ phase plane have shown that there are four basic families of regular orbits together with only one chaotic component. The interesting result is that for large values of the mass of the dark halo the dominant family is the family of the box orbits, while the contribution from the other three families is small. The contribution from the chaotic orbits is also small in this case. Therefore, we can conclude, that in galaxies with massive nuclei and with a massive dark halo, the majority of low angular momentum stars are in box orbits. Furthermore, only a small fraction of stars are in chaotic orbits.

Going a step further, we can say that the dark matter plays an important role, as it affects the regular or chaotic nature of motion as well as the contribution from the different families of regular orbits. Here we must note, that in galaxy models with a spherical dark halo component, in all studied cases, we have found that the percentage of chaotic orbits is small, or negligible, when a massive dark halo is present (see Caranicolas, 1997; Papadopoulos \& Caranicolas, 2006; Caranicolas \& Zotos, 2011). On the contrary, when the dark matter is concentrated in the disk-halo component, it was found that the percentage of the chaotic orbits increases as the amount of dark matter increases in disk galaxies hosting massive nuclei (see Caranicolas, 2012). It is in our future plans to investigate further the role played by the dark matter, not only in relation to the regular or chaotic nature of motion  but also, in relation to the whole set of orbits, using realistic models constructed from recent observational data.

\section*{Acknowledgements}

The authors would like to express their thanks to the anonymous referee for the aptly comments which improved the clarity of the present article.

\end{document}